\documentclass[24pt]{article}
\usepackage{amssymb}
\usepackage{latexsym}
\usepackage{amsmath}
\usepackage[cp1251]{inputenc}
\usepackage{graphicx}

\title{Microscopic theory of heat capacity of liquid helium-4 for temperatures above the critical point}

\author{I.~O.~Vakarchuk, V.~S.~Pastukhov, R.~O.~Prytula}

\begin{document}

\maketitle

\begin{abstract} {In this paper, with the corresponding formula for
 internal energy obtained in Ref. \cite{first},
combined with a simple calculation of the effective mass of
interacting Bose particles, the behavior of the heat capacity of
liquid $^4$He is analyzed numerically for the entire temperature
range. The results agree quite well with experimental data.}

 {\bf Key words}: liquid $^4$He, heat capacity, effective mass

\end{abstract}

\section{Introduction}
Notwithstanding a great number of papers (starting from Refs.
\cite{Bogolubov_47,Bogolubov_Zubarev}) concerned with the
microscopic study of Bose system's properties a good description
of the heat capacity of liquid helium-4 in the whole temperature
range has not yet been created. The first attempts were made by
Brout \cite{brout} where it was shown in the first order of
perturbation theory for the free energy of non-ideal Bose system
that the presence of interaction does change the order of the
phase transition. In \cite{Rovenchak00,Rovenchak_01} the
thermodynamic functions of liquid helium-4 at all temperatures
were obtained using the two-time temperature Green's function
formalism. A good agreement of the specific heat at low
temperatures with experimental data was obtained and the
temperature of phase transition was calculated as $T_c=1.99$ $K$.
The success of such an approach lies in the application of the
experimentally measured structure factor of liquid helium-4
extrapolated to zero temperature \cite{Rovenchak00} instead of the
interparticle interaction potential. A good agreement of the heat
capacity curve with relevant experimental data for the
temperatures below the temperature of the $\lambda$-transition was
obtained in Ref. \cite{first} where the calculations were made
using the quantum-statistical approach based on the density matrix
of Bose liquid. At a higher temperatures the specific heat curve
was shifted upward almost in a parallel way.

In Ref. \cite{Lindenau00} the density matrix formalism with the
functional optimization of the Jastrow wave-function parameters
was used to describe the properties of liquid helium. The results
for internal energy agree well with experimental data for the
temperatures below the critical one. It was shown that by taking
into account the dynamic two-particle correlations only one can
obtain the value of the critical temperature 3.4~K. Thus for good
agreement with experiments one needs to take into account
higher-order approximations, which are specifically related to the
concept of effective mass of the helium atom in a liquid.

In recent years much attention has been paid to the study of the
atom's effective mass in liquid helium because in this way part of
the interaction could be taken into account accordingly to
Feynman's idea \cite{Feynman_53}. However, there is no
satisfactory formula for the effective mass of the helium atom in
the liquid at arbitrary temperatures. Various scholars were mostly
concerned with the value of the effective mass at $T\to0$. Isihara
and Samulski \cite{Samulski} have used the value of $m^*/m=1.71$
to agree the theoretically calculated sound branch of the
excitation spectrum with the corresponding experimental data. In
Ref. \cite{visnyk_93} the effective mass $m^*/m=1.70$ was obtained
on the basis of the liquid helium-4 structure factor measurements.
In Ref. \cite{Rovenchak03} the interatomic potential was preserved
as the input information, but in part the contribution of higher
correlations was ``transferred'' to the kinetic energy term. In
this way the  mass of particles was renormalized that is somehow
in correlation  with the approach of Ref. \cite{Feynman_53}. As a
result of such a renormalization the value of $m^*/m=1.58$ was
obtained using Green's function method. It was shown in Refs.
\cite{Vakar_96} that the above-mentioned mass renormalization
leads to the expressions obtained for the effective mass of the
$^3$He impurity atom in liquid $^4$He but with the replacement of
the ``pure'' $^3$He atom mass by the $^4$He atom mass.

The aim of this paper is to calculate the heat capacity of liquid
helium above the temperature of phase transition. The formula for
the internal energy of Bose liquid, obtained in Ref. \cite{first}
with the help of the method proposed in Ref. \cite{Vakarchuk04}
(where the effective mass is a free parameter of the theory),
forms the basis of these calculations. Thus, further we calculate
step by step with the help of thermodynamic perturbation theory
the quasi-particle spectrum of the Bose system at the temperatures
higher than the critical one, then we obtain the effective mass
and numerically analyze the behavior of the heat capacity.

\section{Perturbation theory for the grand canonical potential at $T>T_c$}
Consider a collection of $N$ spinless particles embedded into
volume $V$. The Hamiltonian of the system which takes into account
only pair interaction between particles may be written using the
secondary quantization language
\begin{eqnarray}
H=H_0+\frac{N^2}{2V}\nu(0)+\frac{1}{2}\sum_{{\bf k}\neq
0}\nu(k)\rho_{\bf k}\rho_{\bf -k}-\frac{N}{2V}\sum_{{\bf k}\neq
0}\nu(k),
\end{eqnarray}
\begin{eqnarray*}
H_0=\sum_{\bf p}(\varepsilon_{p}-\mu)a^{+}_{\bf p}a_{\bf p}.
\end{eqnarray*}
The creation $a^{+}_{\bf p}$ and destruction $a_{\bf p}$ operators
of the particle with the momentum $\hbar\bf p$ satisfy the usual
bosonic commutation relations. The notations $\nu(k)$ stands for
the Fourier transform of the potential and $\varepsilon_p=\hbar^2
p^2/2m$ for the free-particle spectrum are introduced. It is more
convenient to work in the grand canonical ensemble. That is why we
introduced the chemical potential $\mu$ and fugacity $z=e^{\beta
\mu}$($\beta =1/T$, where $T$ is the temperature) of the system
that can be found with the help of the following equation
\begin{eqnarray}
\sum_{\bf p}\langle a^{+}_{\bf p}a_{\bf p}\rangle=N.
\end{eqnarray}
Using the secondary quantization formalism it is easy to write
down the Fourier transform of the particle density fluctuation
operator
\begin{eqnarray}
\rho_{{\bf k}}=\frac{1}{\sqrt{V}}\sum_{\bf p}a^{+}_{\bf p}a_{\bf
p+k}, \ \ ({\bf k}\neq 0).
\end{eqnarray}
Further, our task is to calculate the partition function of a
many-boson system above the temperature of phase transition. Of
cause, the most interesting features of these calculations occur
at the region in a close vicinity of the temperature of Bose
condensation. In the statistical operator, let us pass to the
interaction representation and write down the partition function
in the following way:
\begin{eqnarray}\label{Z}
Z=\textrm{Sp}\left\{e^{-\beta H}\right\}=
Z_0\exp\Bigg\{-\beta\frac{N^2}{2V}\nu(0)+\beta\frac{N}{2V}\sum_{{\bf
k}\neq 0}\nu(k)\Bigg\}\times\nonumber\\
\times\left\langle
T_{\tau}\exp{\Bigg\{}-\frac{1}{2}\int\limits^{\beta}_{0}d\tau
\sum_{{\bf k}\neq 0}\nu(k)\rho_{\bf k}(\tau)\rho_{\bf
-k}(\tau){\Bigg\}}\right\rangle_0,
\end{eqnarray}
where the quantity
\begin{eqnarray*}
\rho_{\bf k}(\tau)=e^{\tau H_0}\rho_{\bf k}e^{-\tau H_0},
\end{eqnarray*}
and the braces stand for statistical averaging with the
Hamiltonian $H_0$. The first multiplier $Z_{0}$ is the partition
function of the ideal Bose gas. The second one and the third one
 take into account the inter-particle interaction completely.

Next, we rewrite, with the help of the Hubbard-Stratonovich
transformation, the $T$-exponent in terms of the functional
integral and we also average it by the states of the ideal Bose
gas
\begin{eqnarray}\label{Z_int}
Z/Z_0&=&\exp\Bigg\{-\beta\frac{N^2}{2V}\nu(0)+\beta\frac{N}{2V}\sum_{{\bf
k}\neq 0}\nu(k)\Bigg\}\nonumber\\
&\times&\int
D\varphi\exp\left\{-\frac{1}{2}\sum_{q}\left(1+\nu(k)\Pi(\omega_n,
k )\right)
\varphi_{q}\varphi_{-q}\right.\nonumber\\
&+&\sum_{l\geq
3}\frac{(-i)^l}{l!(V\beta)^{l/2-1}}\mathop{\sum_{q_1}\ldots
\sum_{q_l}}\limits_{q_1+ \ldots + q_l=0}\sqrt{\nu(k_1)}\ldots
\sqrt{\nu(k_l)}\Pi_l(q_1,\ldots, q_l)\varphi_{q_1} \ldots
\varphi_{q_l}\Bigg\}, \ \ \ \
\end{eqnarray}
where we use the notations $q=(\omega_n,\bf k)$ and $\omega_n=2\pi
n T \ (n=0;\pm 1;\pm 2; \ldots)$ is the Matsubara frequency and
\begin{eqnarray*}
\sum_{q}=\sum_{\omega_n}\sum_{{\bf k}\neq 0}.
\end{eqnarray*}
Here the symbol $\int D\varphi$ denotes the integration over real
and imaginary parts of the $\varphi_q$ variables from the half
space of all possible values of $q$ due to the symmetry
$\varphi^*_q=\varphi_{-q}$. The polarization operator
\begin{eqnarray}\label{Pi}
\Pi(\omega_n, k
)=\frac{1}{V\beta}\sum_{q'}G_0(q')G_0(q'+q)=\frac{1}{V}\sum_{\bf
k'}\frac{n_{{\bf k'}+{\bf k}}- n_{\bf
k'}}{i\omega_n-\varepsilon_{{\bf k'}+{\bf k}}+\varepsilon_{\bf
k'}}.
\end{eqnarray}
Here and thereafter $n_k$ is the filling factor of the ideal Bose
gas. We also introduced notations for the symmetrical functions
\begin{eqnarray}\label{Pi_l}
\Pi_l(q_1,\ldots, q_l)=\frac{1}{V\beta
l}\sum_{q}\big\{G_0(q)G_0(q+q_1) \ldots G_0(q+q_1+\ldots
+q_{l-1})+\textrm{permutations}\big\}.
\end{eqnarray}
The one-particle Green's function of noninteracting bosons is
\begin{eqnarray*}
G_0(q)=\frac{1}{i\omega_n-\varepsilon_{k}+\mu}.
\end{eqnarray*}
The representation Eq. (\ref{Z_int}) when the partition function
is written in terms of functional integrals was used successfully
in the theory of Fermi systems with Coulombic interaction in Ref.
\cite{vavrukh}. The fact that in the Gaussian approximation of the
calculation of the functional integral we recover Random Phase
Approximation (RPA) correctly is a great advantage of our method.
This approximation in the case of interacting fermions generalizes
at finite temperatures the well-known result of
Gell-Mann--Brueckner for the high-density electron gas. The
non-Gaussian part can be taken into account approximately by means
of perturbation theory.

The thermodynamic potential up to the first order of the
perturbation theory (in two-sum approximation over the
``4-vector'') is
\begin{eqnarray}\label{Omega}
\Omega=\Omega_0+\Omega_1+\Omega_2,
\end{eqnarray}
where the ideal gas contribution
\begin{eqnarray}
\Omega_0=T\sum_{\bf p}\ln\left(1-z e^{-\beta
\varepsilon_p}\right),
\end{eqnarray}
and the RPA-part
\begin{eqnarray}\label{Omega_1}
\Omega_1=\frac{N^2}{2V}\nu(0)
+\frac{1}{2\beta}\sum_{q}\ln\left|1+\nu(k)\Pi(\omega_n, k )\right|
-\frac{N}{2V}\sum_{{\bf k}\neq 0} \nu(k).
\end{eqnarray}
As we work in the grand canonical ensemble we take the average
number of particles $N$ of the function of the chemical potential
$\mu$.

The one-loop contribution to the thermodynamic potential is
\begin{eqnarray}\label{Omega_2}
\Omega_2&=& \frac{1}{2\cdot3!V\beta^2}\mathop{\sum_{q_1}\sum_{q_2}
\sum_{q_2}}\limits_{q_1+ q_2 + q_3=0}\Pi_3(q_1,q_2,
q_3)\Pi_3(-q_1,-q_2,
-q_3)f(q_1)f(q_2)f(q_3)\nonumber \\
&-&\frac{1}{8V\beta^2}\sum_{q_1}\sum_{q_2}\Pi_4(q_1,-q_1, q_2,
-q_2)f(q_1)f(q_2).
\end{eqnarray}
The function $f(q)=\nu(k)\langle\varphi_{q}\varphi_{-q}\rangle$,
and the correlator
\begin{eqnarray*}
\langle\varphi_{q}\varphi_{-q}\rangle=\frac{1}{1+\nu(k)\Pi(\omega_n,
k )}
\end{eqnarray*}
were obtained in the Random Phase Approximation. The structure of
expressions (\ref{Omega_1}), (\ref{Omega_2}) clearly shows that
they are ``in correlation'' with the formulae obtained in Ref.
\cite{first} where an entirely different method of calculations
was used.

\section{Renormalization of the one-particle spectrum}

It is clear that the basis of further calculations is fully
determined by the renormalization of the quasi-particle spectrum.
For our analysis we use RPA. Notwithstanding the simplicity of
this approximation it ``catches'' certain important features of
the behavior of the system. It is not surprising because RPA
effectively sums up an infinite set of terms of the perturbation
theory divergent near phase transition point.

At first let us use the thermodynamic equality $-\partial
\Omega/\partial \mu=N$ to find the average number of particles in
the system. The explicit calculation of the corresponding
derivative with the first two terms of Eq. (\ref{Omega}) gives
\begin{eqnarray}\label{N}
N&=&\sum_{\bf p}\bigg\{n_p-\frac{N}{V}\nu(0)\frac{\partial
n_p}{\partial \mu}+ \frac{1}{2V}\sum_{{\bf k}\neq
0}\nu(k)\frac{\partial n_p}{\partial
\mu}\nonumber\\
&-&\frac{1}{2V\beta}\sum_{q}\frac{\nu(k)}{1+\nu(k)\Pi(\omega_n,k)}
\left[\frac{1}{\varepsilon_{|{\bf k}-{\bf p}|}-\varepsilon_{p}-i
\omega_n}+(\omega_n\rightarrow -\omega_n)\right]\frac{\partial
n_p}{\partial \mu}\bigg\}.\qquad
\end{eqnarray}
Let us construct the Bose filling factor with a new spectrum using
the expression in braces. Making use of equality $\partial
n_p/\partial \mu=-\partial n_p/\partial \varepsilon_p$ we finally
obtained the formula for the renormalized one-particle spectrum
\begin{eqnarray}
{\varepsilon}^*_{p}=\varepsilon_{p}+ \Delta{\varepsilon}_{p}, \ \
{\mu}^*=\mu+\Delta \mu,
\end{eqnarray}
where the correction to the quasi-particle spectrum is
\begin{eqnarray}\label{Delta_xi}
\Delta{\varepsilon}_{p}=\frac{1}{\beta V}\sum_{q}\frac{
\nu(k)}{1+\nu(k)\Pi(\omega_n,k)}
\left\{\frac{1}{\varepsilon_{|{\bf k}-{\bf p}|}-\varepsilon_{p}-i
\omega_n}-\frac{1}{\varepsilon_{k}-i \omega_n} \right\},
\end{eqnarray}
and the correction to the chemical potential is
\begin{eqnarray}\label{Delta_mu}
\Delta \mu=-\frac{N}{V}\nu(0)- \frac{1}{\beta V}\sum_{q}\frac{
\nu(k)}{1+\nu(k)\Pi(\omega_n,k)}\frac{\varepsilon_{k}}{
\varepsilon^2_{k}+\omega^2_n}+\frac{1}{2V}\sum_{{\bf k}\neq
0}\nu(k),
\end{eqnarray}
It is easy to obtain the above-mentioned expression for the
spectrum in a different way. To do this one has to recall that
variational derivative of the $\Omega$-potential with respect to
$\varepsilon_p$ equals the renormalized one-particle filling
factor ${n}^*_p$. After simple calculations we obtained the
following formula:
\begin{eqnarray}\label{tild_n_p}
{n}^*_p&=&n_p+\frac{N}{V}\nu(0)\frac{\partial n_p}{\partial
\varepsilon_p}-\frac{1}{2V}\sum_{{\bf k}\neq
0}\nu(k)\frac{\partial n_p}{\partial
\varepsilon_p}\nonumber\\
&+&\frac{1}{2V\beta}\sum_{q}\frac{\nu(k)}{1+\nu(k)\Pi(\omega_n,k)}\nonumber\\
&\times&\bigg\{\left[\frac{1}{\varepsilon_{|{\bf k}-{\bf
p}|}-\varepsilon_{p}-i \omega_n}+(\omega_n\rightarrow
-\omega_n)\right]\frac{\partial n_p}{\partial
\varepsilon_p}\nonumber\\
&+&\left[\frac{n_p-n_{|{\bf k}-{\bf p}|}}{(\varepsilon_{|{\bf
k}-{\bf p}|}-\varepsilon_{p}-i \omega_n)^2}+(\omega_n\rightarrow
-\omega_n)\right]\bigg\}.
\end{eqnarray}
It is easy to argue by making summation over the wave-vector $\bf
p$ of the left-hand and right-hand sides of the previous equality
that the second term in braces vanishes. So, after the summation
of expression (\ref{tild_n_p}) we arrive at equality (\ref{N}) and
thus we get the same expression (\ref{Delta_xi}) for the
correction to a one-particle spectrum again. It is interesting to
note that the calculation of the variation derivative
$\delta\Omega/\delta n_p$ in RPA gives the same result. Finally,
Eq. (\ref{Delta_xi}) coincides with the result derived in Ref.
\cite{past} where calculations were made in terms of temperature
Green's function technique.

Let us analyze expression (\ref{Delta_xi}). First, the potential
problems with the integration over the wave-vector may occur only
in the critical region and at the zero frequency $\omega_n$. That
is why we write down this term apart and immediately set apart the
Hartree-Fock-like term
\begin{eqnarray}\label{Delta_xi_2}
\Delta{\varepsilon}_{p}&=&\frac{1}{\beta V}\sum_{{\bf k}\neq
0}\frac{ \nu(k)}{1+\nu(k)\Pi(k)}\left[\frac{1}{ \varepsilon_{|{\bf
k}-{\bf
p}|}-\varepsilon_{p}}-\frac{1}{ \varepsilon_k}\right]\nonumber\\
&+&\frac{1}{V}\sum_{{\bf k}\neq
0}\nu(k)\left\{n(\beta\varepsilon_{|{\bf k}-{\bf
p}|}-\beta\varepsilon_{p})-n(\beta\varepsilon_k)-\frac{1}{\beta[\varepsilon_{|{\bf
k}-{\bf p}|}-\varepsilon_{p}]}+\frac{1}{\beta\varepsilon_k}\right\}\nonumber\\
&-& \frac{1}{\beta V}\sum_{q \neq 0}\frac{
\nu^2(k)\Pi(\omega_n,k)}{1+\nu(k)\Pi(\omega_n,k)}\left[
\frac{1}{\varepsilon_{|{\bf k}-{\bf p}|}-\varepsilon_{p}-i
\omega_n}-\frac{1}{\varepsilon_k-i \omega_n}\right],
\end{eqnarray}
where $\Pi(k)\equiv \Pi(0, k)$, $n(x)=1/(e^x-1)$ and  $\rho=N/V$
is equilibrium density of the system. Secondly, to go further we
have to investigate the properties of polarization operator
(\ref{Pi})
\begin{eqnarray}\label{Pi_int}
\Pi(\omega_n, k)&=&\frac{\beta
k^3_0}{(2\pi)^2}\frac{k_0}{2k}\int\limits^{\,\infty}_{0}dx\,
\frac{x }{z^{-1}e^{x^2}-1}\nonumber\\
&\times&\left\{
\ln\left|\frac{(k/k_0)^2+2xk/k_0-iu_n}{(k/k_0)^2-2xk/k_0-iu_n}
\right|+(u_n\rightarrow -u_n)\right\},
\end{eqnarray}
here for convenience the following notations are used:
$k_0=\sqrt{2mT}/\hbar$, $u_n =2\pi n$. We are interested in a long
wave-length behavior of the polarization operator. To find the
leading order asymptote of $\Pi(\omega_n, k)$ at zero frequency in
the critical point it is sufficient to replace the Bose filing
factor $1/(e^{x^2}-1)$ in integral (\ref{Pi_int}) by $1/x^2$. Then
after a simple integral calculation we get $\Pi(k \rightarrow 0)=
\beta k^4_0/(8 k)$. For higher temperatures ($\mu\neq 0$)
\begin{eqnarray}\label{Pi-simpl}
&&\Pi(k)=\beta\rho\left\{g_{1/2}(e^{\beta\mu})-\frac{1}{6}\frac{k^2}{k^2_0}g_{-1/2}(e^{\beta\mu})\right\}{\bigg/}g_{3/2}(e^{\beta\mu})
+\ldots,\nonumber\\
&&g_{l}(e^{y})=\sum_{n\geq1}\frac{e^{yn}}{n^{l}}.
\end{eqnarray}
The dots stand for higher than quadratic terms in the expansion
over the wave-vector. It is easy to see from definition
$(\ref{Pi})$ that for non-zero frequencies $\Pi(\omega_n,0)=0$.

Now, let us consider the contribution to the one-particle spectrum
(\ref{Delta_xi}) from zero frequency. Using designation $\Delta
\varepsilon^{u}_p$ for this term of spectrum it is easy to write
down
\begin{eqnarray}\label{Delta_d}
\Delta \varepsilon^{u}_p=(k_0/\pi)^2p\int\limits^{\,\infty}_{0}dx
\frac{\nu(2xp)}{1+\rho \nu(2xp)\Pi(2xp)} \left[\frac{x}{2}
\ln{\Big|}\frac{x+1}{x-1}{\Big|}-1\right].
\end{eqnarray}
For self-consistency of our calculations, especially near the
critical point, the chemical potential $\mu$ should be changed by
$\mu^*$ in the right-hand side of Eq. (\ref{Delta_d}) (the
critical point is determined by the equation $\mu^*=0$,
respectively). Admittedly, the ideal gas dispersion relation
should be replaced by the exact one-particle spectrum, but further
analysis will not be influenced by this replacement qualitatively.

Let us consider the value of the integral in Eq. (\ref{Delta_d})
at a small $p$ and assume for definiteness that the temperature is
higher than the critical one. Then substituting $\Pi(2xp)
\rightarrow \Pi(0)$, and $\nu(2xp) \rightarrow \nu(0)$ we obtain
\begin{eqnarray*}
\Delta \varepsilon^{u}_p=(k_0/\pi)^2p\frac{\nu(0)}{1+\rho
\nu(0)\Pi(0)}\int\limits^{\,\infty}_{0}dx
 \left[\frac{x}{2}
\ln{\Big|}\frac{x+1}{x-1}{\Big|}-1\right].
\end{eqnarray*}
It turns out that this integral equals zero identically. Moreover
even after the substitution of $\Pi(k)\rightarrow$
(\ref{Pi-simpl}), the integral in Eq. (\ref{Delta_d}) equals zero
too. Thus it is shown, with a realistic restriction on the Fourier
transform of the potential energy, i.e. the absence of linear and
quadratic terms in the expansion of $\nu(k)$ at a small $k$, that
$\Delta \varepsilon^{u}_p= o(p^2)$.

The situation is quite different in the critical region. Here the
leading order asymptote is $\Delta \varepsilon^{u}_p \sim
p^2\ln(p)$ (it is not hard to ascertain taking into account the
properties of $\Pi(k)\sim {1/k}$ in this region), which obviously
is a hint at the following behavior of the one-particle spectrum
$\Delta \varepsilon^{u}_p \sim p^{2-\eta}$ ($\eta\rightarrow 0$)
at the critical temperature. Clearly one cannot obtain this result
using a simple perturbative approach.

Hence, having separated the non-analytic problematic part of the
spectrum (the second term in Eq. (\ref{Delta_xi_2})) we can
consider the ``non-universal'' one, i.e. the remainder of Eq.
(\ref{Delta_xi_2}). Precisely this expression will determine the
observed non-universal properties of the Bose liquid. The latter
calculations are linked to the summation over the Matsubara
frequency in the last term of Eq. (\ref{Delta_xi_2}) and coincide
with those in Ref. \cite{past}. That is why we do not dwell on the
details of these calculations. Now the leading-order non-vanishing
term of the quasi-particle dispersion relation is quadratic over
the wave-vector. We recall that $\Delta \varepsilon^{u}_p= o(p^2)$
and hence its contribution is not significant. So, for reasons of
simplification we assume the spectrum to be a quadratic
free-particle one, but with the normalized mass.

As we single out the ``problematic'' contribution to the
quasi-particle spectrum expanding second and third sums of Eq.
(\ref{Delta_xi_2}) into a series in $p$ we obtain for the
one-particle spectrum
\begin{eqnarray}
\varepsilon^*_p=\Delta \varepsilon^{u}_p+\frac{\hbar^2p^2}{2m^*}.
\end{eqnarray}
The effective mass is
\begin{eqnarray}\label{mass}
m/m^*=1-\Delta(T),
\end{eqnarray}
where the quantity
\begin{eqnarray}\label{Delt}
\Delta(T)&=&\frac{1}{3N}\sum_{{\bf k} \neq
0}\frac{(\alpha_k-1)^2}{\alpha_k(\alpha_k+1)}\nonumber\\
&+&\frac{2}{3N}\sum_{{\bf k} \neq 0} \Big\{
\frac{\alpha^2_k+3}{\alpha^2_k-1}\left[n(\beta
\varepsilon_k)-1/\beta \varepsilon_k\right]\nonumber\\
&-&\frac{3\alpha^2_k+1}{\alpha_k(\alpha^2_k-1)}
\left[n(\beta E_k)-1/\beta E_k\right]\nonumber\\
&+&2\left[1/\beta \varepsilon_k-\beta \varepsilon_k n(\beta
\varepsilon_k)[1+n(\beta \varepsilon_k)]\right]\Big\}.
\end{eqnarray}
Here we use the following notations: $E_k=\alpha_k \varepsilon_k$
is Bogoliubov spectrum, and $\alpha_k=\sqrt{1+2\rho
\nu(k)/\varepsilon_k}$. The effective mass in the low-temperature
region is always larger than its ``bare'' one which means that the
renormalized temperature of the Bose condensation of interacting
particles is always lower than the critical temperature of the
ideal gas. This is the most important result of Eqs. (\ref{mass})
and (\ref{Delt}). At low temperatures carefully calculating the
limit of $\beta \rightarrow 0$  it is easy to ascertain that the
effective mass tends to the mass of particles. It is important
that the temperature-independent part of formula (\ref{Delt})
coincides with the effective mass derived in Ref. \cite{Vakar_96}
where a different method was used for the calculations.

At the end of this section one more remark on the applicability of
the formula for $\Delta(T)$ has to be made. The approximation of
the exact spectrum of collective modes by the Bogoliubov spectrum
on the one hand made it possible to obtain the analytical
expression (\ref{Delt}), on the other hand it brought us beyond
the limits of RPA. It is hard to assess the accuracy of such a
trick, but it can be justified only considering that the Fourier
transform of a two-particle potential is a rapidly decreasing
function. Then the main contribution to the integral over the
wave-vector comes from a lower limit of integration where the
Bogoliubov spectrum completely coincides with the exact one.

\section{Internal energy and heat capacity}

The expression for the grand canonical potential derived in the
second section of this paper is applicable only for the
temperatures higher than the critical one. To describe the
$\lambda$-transition phenomenon, in particular thermodynamic
functions, let us use an approach based on the density matrix of
the Bose liquid \cite{Vakarchuk04}. Using this approach the
dependence of an internal energy of the Bose liquid on the
effective mass of the helium-4 atom was found in the approximation
of pair-particle correlations in Ref. \cite{first}. In the case of
${m^*/ m=1}$ in Ref. \cite{first} a good agreement of the heat
capacity with experimental data in the region below the critical
point was obtained, but above the temperature of phase transition
the heat capacity curve was shifted upward. Hence, the calculation
of heat capacity for the case of ${m^*/ m\neq1}$ is an interesting
problem.

We take the expression for the internal energy in the
approximation of pair-particle correlations from Ref.
\cite{first}:
\begin{eqnarray} \label{E}
E&=&N{mc^2\over{2}}+\sum_{{\bf k}\neq0}{\hbar^2k^2\over2m}
{1\over{z_0^{-1}e^{\beta\varepsilon^*_k}-1}} +{1\over2}{m^*\over
m}\sum_{{\bf k}\neq0} {\lambda _k\over 1+\lambda_k S_0(k)}
 {\partial S_0(k)\over\partial\beta}\nonumber\\
&+&{1\over4}\sum_{{\bf k}\neq0}
{\hbar^2k^2\over2m}\left(\lambda_k^2+\alpha_k^2-1\right){S(k)}
+{1\over2}\sum_{{\bf k}\neq0}{\hbar^2k^2\over2m}
\left[{\alpha_k\over\sinh\left(\beta E_k \right)}-{1\over\sinh\left(\beta\varepsilon^*_k\right)}\right]\nonumber\\
&+&{1\over 16}\sum_{{\bf k}\neq
0}{\hbar^2k^2\over2m}\left(1-{1\over\alpha_k^2}\right)\left(\alpha_k-{1\over\alpha_k}-4\alpha_k^2\right).
\end{eqnarray}
Here the following notations are introduced:
\begin{eqnarray}\label{struc fact ideal}
S_0(k)=1+2{n_0\over N}n^*_k+{1\over N}\mathop {\sum_{{\bf
p}\neq0}}_{{\bf p}+{\bf k}\neq0}n^*_pn^*_{|{{\bf p}+{\bf k}}|},
\end{eqnarray}
is the structure factor of the ideal Bose gas with a renormalized
mass.

The quantity $n_0$ is an average number of particles of the ideal
Bose gas with zero momentum and $z_0$ is its fugacity.

The next quantity
\begin{eqnarray}
S(k)={S_0(k)\over {1+\lambda_kS_0(k)}} \label{Sq0}
\end{eqnarray}
is the pair structure factor of the Bose liquid, and
\begin{eqnarray}\label{lq}
\lambda_k=\alpha_k\tanh\left({\beta E_k/2}\right)-
\tanh\left({\beta \varepsilon^*_k/2}\right).
\end{eqnarray}

Obviously, if we turn off interparticle interaction $\alpha_k=1$,
$m^*=m$ and take into account that sound velocity in the ideal
Bose gas at $T=0$ equals zero $c=0$ (there are no zero density
fluctuations in the Bose system) and $\lambda_k=0$ (\ref{lq}) we
obtain a well-known formula for the energy of the ideal Bose gas
\begin{eqnarray}
E=\sum_{{\bf k}\neq0}{\hbar^2k^2\over2m}
{1\over{z_0^{-1}e^{\beta\hbar^2k^2/2m}-1}}.
\end{eqnarray}

Let us analyze the total energy Eq.(\ref{E}) in the low
temperature region where it coincides with the formulas obtained
in Ref. \cite{Vakarchuk04,Vakarchuk_II}. At low temperatures
$T\to0$, when only small values of the wave-vector $\bf k$ are
important in the expression for the spectrum $E_k$ we obtain:
\begin{eqnarray}
E=E_0+{V\pi^2\over 30 (\hbar c)^3}T^4\nonumber,
\end{eqnarray}
where the ground state energy is
\begin{eqnarray*}
E_0=N{mc^2\over{2}}-{1\over4}\sum_{{\bf k}\neq0}
{\hbar^2k^2\over2m}\left(\alpha_k-1\right)^{2} +{1\over
16}\sum_{{\bf k}\neq
0}{\hbar^2k^2\over2m}\frac{1}{\alpha_k}\left(\alpha_k-{1\over\alpha_k}\right)^2,
\end{eqnarray*}
 and the heat capacity,
respectively,
\begin{eqnarray}
C_V=V{2\pi^2\over 15(\hbar c)^3}T^3 \nonumber.
\end{eqnarray}

 Exactly the same temperature dependence
of the heat capacity of liquid $^4$He at $T\to0$ is observed.
Since we obtained the correct behavior of the heat capacity at low
temperatures using the energy of the Bose liquid Eq. (\ref{E}) we
expect to derive the correct behavior of the heat capacity for the
entire temperature range.

\section{Numerical results}
Our numerical calculations are carried out at the equilibrium
density of liquid helium $\rho=0.02185$~\AA$^{-3}$, mass of
particles $m=4.0026$~u, sound velocity $c=238.2$ m/s in the limit
of $T\to0$ \cite{Donnelly98}, and at the critical temperature of
the ideal gas $T_c=3.138$~K. We use the liquid structure factor
extrapolated to $T=0$ \cite{Rovenchak00} as the output
information, instead of the interparticle potential, i.e.
\begin{eqnarray}
\alpha_k=\frac{1}{S^{exp}(k)},
\end{eqnarray}
where $S^{exp}(k)$ is the experimentally measured structure factor
at $T=0$.

It is logical to start calculations with the formula for the
renormalized one-particle spectrum and thus with the formula for
the effective mass of particles. Despite the complexity of the
last sum over the wave-vector in Eq. (\ref{Delt}) the main
contribution to the effective mass arises from the
temperature-independent part. In Fig. 1 the dependence of a
dimensionless value of the effective mass $m^*/m$ as a function of
temperature is presented. Formally we extrapolated a curve of the
effective mass in the condensate region where obviously it becomes
a parameter of the theory. It is important that at zero
temperature $m^*$ coincides with the effective mass of the
impurity atom in the Bose liquid \cite{Vakar_96}.

The first three terms of a low-temperature expansion are ($T\ll
mc^2$, $T\ll \hbar^2\rho^{2/3}/m$)
\begin{eqnarray}
\Delta(T\rightarrow
0)=\Delta_0+\Delta_1T-\Delta_{3/2}T^{3/2}+o(T^{5/2}).
\end{eqnarray}
where $\Delta$-coefficients are
\begin{eqnarray*}
\Delta_0=\frac{1}{3N}\sum_{{\bf k} \neq
0}\frac{(\alpha_k-1)^2}{\alpha_k(\alpha_k+1)}=0.41,
\end{eqnarray*}
\begin{eqnarray*}
\Delta_1=\frac{2}{3N}\sum_{{\bf k} \neq
0}\frac{\alpha^2_k-1}{\alpha^2_k\varepsilon_k}=0.31,
\end{eqnarray*}
\begin{eqnarray*}
\Delta_{3/2}=\frac{4}{3}\zeta(3/2)\left(\frac{m}{2\pi\hbar^2}\right)^{3/2}/\rho=0.24.
\end{eqnarray*}
Then the effective mass equals approximately
\begin{eqnarray*}
m^*=m\Big/\left(0.59-0.31T+0.24T^{3/2}\right).
\end{eqnarray*}
This formula reproduces the curve in Fig. 1 quite well up to the
critical temperature.
\begin{figure}[htb!]
\centerline{\includegraphics
[width=0.7\textwidth,clip,angle=-0]{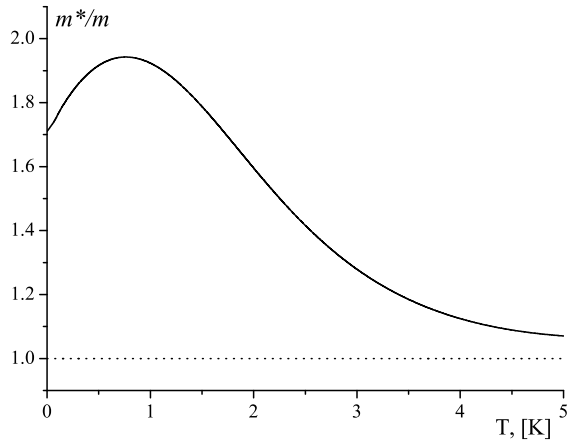}}
\caption{Temperature dependence of the fraction $m^*/m$}
\end{figure}

Now we are in position to calculate the renormalized temperature
of the Bose condensation. We can find $T_c$ using condition
$\sum_{\bf p}n^*_p=N$ at the zero value of the renormalized
chemical potential $\mu^*$. A simple calculation gives $T_c=1.94$
$K$ that agrees quite well with experimental measurements of the
temperature of the $\lambda$-transition $T^{exp}_c=2.17$ $K$
despite the simplicity of the approximations.

Let us pass on to the heat capacity calculation:
\begin{eqnarray}
C_V=\left(\partial E\over\partial T\right)_V.
\end{eqnarray}

We calculate heat capacity using the difference method and build
the plot of its temperature dependence $T/T_c$.

A comparison of different heat capacity curves is depicted in Fig.
2. As is seen from the comparison of the calculated curve 1 with
the experimental one the agreement is quite good at low
temperatures $(0<T/T_c<1)$. At the temperatures $T/T_c>1$ the
inconsistency occurs: the behavior of the calculated heat capacity
is very similar to the behavior of the experimental curve, but
shifted upward almost in a parallel way. This inconsistency is
related to the fact that three- and four-particle correlations
should be taken into account for the quantitative description. The
contribution of three- and four-particle correlations, as is shown
in Refs. \cite{Vakarchuk_II,Hlushak}, improves significantly the
ground-state results and gives a fairly good agreement at $T\to0$.

\begin{figure}[!!h]
\centerline{\includegraphics[scale=1.0,clip]{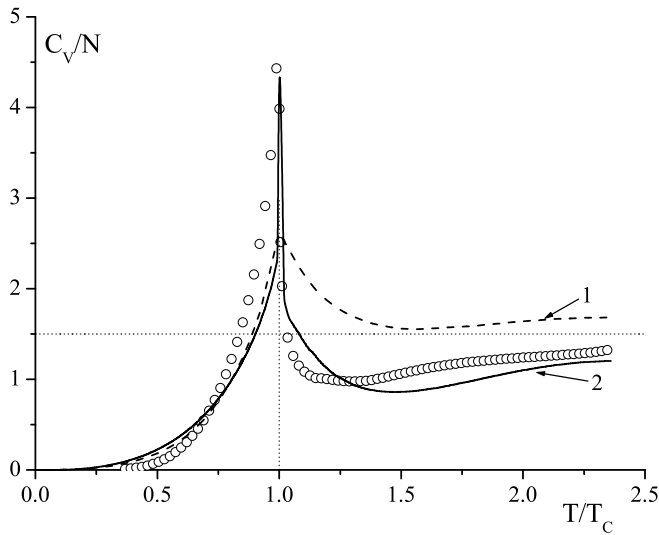}}
\caption{Heat capacity of liquid helium-4. Curve 1 is the
calculated heat capacity in the pair-correlation approximation
(\ref{E}) \cite{first};  curve 2 is the heat capacity in
pair-correlation approximation (\ref{E}) taking into account the
effective mass of the helium atom in liquid (\ref{mass}). The
circles show experimental data from Refs.
\cite{Arp05,Ceperley95,Arp98}.}
\end{figure}

Further, let us calculate the heat capacity of liquid helium-4
with taking into account the effective mass of the helium atom in
the liquid (curve 2). At low temperatures the heat capacity with
taking into account the effective mass of the Bose particles
practically coincides  with curve 1, which agrees well with
experimental data. This shows a weak dependence of the heat
capacity on the effective mass below the temperature of phase
transition. As is seen from Fig. 2, the calculated curve 2 (unlike
curve 1) agrees quite well with the experimental one. It is
related to the fact that by using the effective mass we partially
take into account a contribution from three- and four-particle
correlations. It is not surprising that in close vicinity of the
Bose condensation point the theoretically calculated heat capacity
deviates most significantly from the experimental curve. It is
solely related to the inconsistency of our description near the
critical point because the non-analytical part (\ref{Delta_d}) of
the one-particle spectrum, which makes a significant contribution
in the thermodynamic functions at $T\rightarrow T_c$, is
disregarded in our approach. One has to use renormalization group
methods \cite{kaup,camp,pogorelov} for the correct description of
the heat capacity in this temperature region.

\section{Conclusions}
In this paper we succeed in deriving quite well an agreement of
the heat capacity curve of liquid helium  with experimental data
practically for all temperatures. The application of the
thermodynamic functions of the Bose liquid obtained with the help
of the Hamiltonian averaging combined with a one-particle spectrum
derivation were the key moments of the calculations.
Notwithstanding the simplicity of the spectrum calculation we
obtained quite interesting results. In particular, we can
decompose the part of the quasiparticle spectrum that is
responsible for the non-analyticity in the Bose condensation point
and show that this term of spectrum has no effect on the physical
observables in the undercritical temperature region. So, an
attempt is made to justify microscopically the idea that the
$\lambda$-transition in a real quantum liquid is very similar to
the Bose-Einstein condensation phenomena of the ideal gas
``slightly'' deformed by the interaction between the particles
(keeping in mind the non-universal properties of the system).

 The calculation found that the long wave-length limit of
the ``non-universal'' part of the one-particle spectrum is
quadratic over the wave-vector, i.e. very similar to the
dispersion relation of the ideal gas but with a new mass. In the
general case it is shown that this new mass at low temperatures is
always greater than the mass of particles, and thus, the presence
of the interaction at least in our approximation always lowers the
critical temperature.

Another feature of the developed theory, perhaps a bit unexpected,
is that even this simple temperature dependence of the effective
mass improves the behavior of the heat capacity curve in the
undercritical region and does not affect it in the condensate
phase. Hence, the quantum-statistical approach based on the
density matrix is suitable for describing thermodynamic properties
of such a strongly-interacting Bose liquid as the helium-4 liquid
not only in the limits of low and high temperatures, but for the
entire temperature range.

\end{document}